\documentclass[prl,superscriptaddress,showpacs,twocolumn]{revtex4}
\usepackage{graphicx}

\begin{document}
\title{Specific Heat of the Quantum Bragg Glass}
\author{Gregory Schehr}
\affiliation{LPTENS CNRS UMR 8549 24, Rue Lhomond 75231 Paris
Cedex 05, France}
\author{Thierry Giamarchi}
\affiliation{DPMC, University of Geneva, 24 Quai Ernest-Ansermet,
CH-1211 Geneva, Switzerland}
\author{Pierre Le Doussal}
\affiliation{LPTENS CNRS UMR 8549 24, Rue Lhomond 75231 Paris
Cedex 05, France}
\date{\today}

\begin{abstract}
We study the thermodynamics of the vibrational modes
of a lattice pinned by impurity disorder
in the absence of topological defects (Bragg glass
phase). Using a replica variational method we compute the specific heat $C_v$
in the quantum regime and find $C_v \propto T^3$
at low temperatures in dimension three and two.
The prefactor is controlled by the pinning
length. The non trivial cancellation of the
linear term in $C_v$ arises from the so-called marginality
condition and has important consequences for other
mean field models.
\end{abstract}
\pacs{}
\maketitle

Understanding the behavior of the specific heat in disordered and
glassy systems
is a longstanding theoretical challenge \cite{zeller_chalspe_struct_glasses}.
Its low temperature dependence has been observed to be linear
in a variety of glasses, including amorphous solids, disordered
crystals and spin glasses
\cite{phillips_chalspe_amorphous,ackerman_chalspe_disordered_crystals,binder_spinglass_review}.
A phenomenological interpretation was proposed
{\it assuming} the existence of two level systems
\cite{anderson_twolevels} leading to such a linear behavior.
Despite its remarkable success for many systems the
range of applicability and microscopic justification of such arguments
are still open questions. Analytical calculations
based on models of quantum solids with structural
disorder account for higher frequency features such as
the ``boson peak''  \cite{boson_peak_review,
schirmacher_bosonpeak,grigera_vibrational_spectrum}
but produce only phonon-like specific heat $C_v \sim T^3$, except in the classical
limit \cite{schehr_chalspe_classique,schirmacher_anharmonic}. On the other hand
new possibility arises from recent developments of the mean field methods
\cite{mezard_variational_replica,giamarchi_columnar_variat}
to quantum spin glasses
\cite{georges_mf_quantum_spinglass,cugliandolo_quantum_p_spin}.
The glass phase is there described by a replica symmetry broken saddle
point solution and evidence was found
\cite{georges_mf_quantum_spinglass,cugliandolo_quantum_p_spin}
for a linear specific heat
both in the $p$-spin and Heisenberg spin glasses. Since even
in these solvable models the specific heat remains delicate
to extract analytically, a later numerical study claims
instead $C_v \sim T^2$ \cite{rozenberg_t2} and compares to
experiments. It is thus still an outstanding problem
to understand the behavior of the specific heat
already at the level of the mean field description
of quantum glasses.

In addition to structural and spin glasses, there has been considerable
recent interest in pinned elastic systems where disorder originates
from substrate impurities. Such systems cover
a wide range of experimental situations both in the classical and
quantum limit, such as vortex lattices in superconductors
\cite{blatter_vortex_review,nattermann_vortex_review,giamarchi_book_young,giamarchi_vortex_review},
electron crystals \cite{giamarchi_wigner_review,chitra_wigner_long}, charge and spin
density waves \cite{gruner_book_cdw}, disordered liquid crystals \cite{saunders_smectic_bragg_glass},
etc.. All these systems are characterized by a competition between
disorder and elasticity, which leads to pinning and glassy
behavior. Although the question of positional order and
correlations is now better understood, with reasonable
agreement between theory and experiments, the behavior
of the specific heat is still largely not understood
and, in some cases, affected by non equilibration effects
\cite{odin_chalspe_blue_bronze,lasjaunias_chalspe_noneq_sdw,blatter_chalspe,
izawa_chalspe_borocabide,bulaevskii_chalspe,
caroli_core_states,volovik_chalspe_dwaves}.

It is thus crucial to develop a first principle method
to compute the temperature dependence of the specific heat in such disordered
elastic systems. Here we address this question for a Bragg glass
system allowing both for thermal and quantum fluctuations
\cite{giamarchi_columnar_variat} and obtain a quite general
formula to compute the specific heat. We find that the term
proportional to $T$ in
the specific heat vanishes. This leads to a $T^3$ behavior of $C_v$
in all dimensions between $2$ and $4$. The
Larkin Ovchinnikov pinning length is found to control the
coefficient. The result holds for a periodic object, i.e. a Bragg
glass, and for interfaces with continuous degrees of freedom.
We elucidate the general mechanism leading, in
the mean field approach, to the cancellation of the
term linear in $T$. We discuss how this property
allows to extract such a term in related quantum spin glass models.
In a companion paper \cite{schehr_chalspe_classique} we investigate the
classical limit and the applications to vortex lattices.

Let us consider a collection of interacting quantum objects of
mass $\mu$ whose position variables are denoted by a $m$-component
vector
$u_\alpha(R_i,\tau)$. The equilibrium positions $R_i$ in the
absence of any fluctuations form a perfect lattice of spacing $a$.
Interactions result in an elastic tensor $\Phi_{\alpha,\beta}(q)$
which describes the energy associated to small displacements, the
phonon degrees of freedom. Impurity disorder is modeled by a
$\tau$ (imaginary time) independent gaussian random potential $U(x)$ interacting
with the local density $\rho(x) = \sum_i \delta (x - R_i -
u(R_i,\tau))$. We will describe systems in the weak disorder
regime $a/R_a \ll 1$ where $R_a$ is the translational correlation
length, e.g. in a Bragg glass phase where the condition
$|u_\alpha(R_i,\tau) - u_\alpha(R_i+a ,\tau)| \ll a$ holds, no
dislocations being present.
At equilibrium the system is described by the partition function $Z=
\text{Tr} e^{- \beta H} = \int Du D\Pi e^{-S/\hbar}$ with Hamiltonian
$H=H_{\text{ph}} + H_{\text{dis}}$:
\begin{eqnarray}
&& H_{\text{ph}} = \frac{1}{2} \int_q \frac{\Pi(q)^2}{\mu} +
\sum_{\alpha,\beta}
u_{\alpha}(q)
\Phi_{\alpha,\beta}(q) u_{\beta}(-q)  \nonumber \\
&& H_{\text{dis}} = \int d^dx U(x) \rho(x, u(x)) \label{Hsys}
\end{eqnarray}
and its associated Euclidean quantum action $-S[u,\Pi]=\int_\tau
\int_q i \Pi_{\alpha}(q) \partial_\tau u_{\alpha}(q) - H$. Here
$\int_q \equiv \int_{BZ} \frac{d^d
q}{(2 \pi)^d}$ denotes integration on the first Brillouin zone and
all integrals over the imaginary time variable $\tau$ go from $0$
to $\beta \hbar$, $\beta=1/T$ being the inverse temperature.

For simplicity we illustrate the calculation on a
isotropic system with $\Phi_{\alpha,\beta}(q)= c q^2 \delta_{\alpha
\beta}$ and denote disorder correlations $\overline{U(x) U(x')} =
\Delta(x-x')$.
The disorder average is performed using the replica trick
$\overline{Z^k}=\int Du e^{-S_{\text{eff}}/\hbar}$ and integrating over
$\Pi$, after some manipulations \cite{giamarchi_columnar_variat}
one obtains the following replicated action $S_{\text{eff}} =
S_{\text{ph}} + S_{\text{dis}}$ with:
\begin{eqnarray}
&& S_{\text{ph}} = \int d^d x d\tau \frac{c}{2} \sum_a (\nabla_x
u_a)^2 + \frac{1}{v^2} (\partial_{\tau} u_a)^2  \nonumber \\
&& S_{\text{dis}}=-  \frac{1}{2 \hbar} \int d^d x d\tau d\tau'
\sum_{ab} R(u_a(x,\tau) - u_b(x,\tau')) \nonumber \\
&& R(u) = \rho_0^2 \sum_K \Delta_K \cos(K \cdot u)
\label{Srepliquee}
\end{eqnarray}
Here
$v=\sqrt{c/\mu}$ is the pure phonon velocity and
$\Delta_K = \int d^d x e^{i K \cdot x} \Delta(x)$ denote the
harmonics of the disorder correlator at the reciprocal lattice
vectors $K$, and $\rho_0 \sim a^{-2}$ the average areal density.

This quantum model is then studied using the Gaussian Variational
Method (GVM) on
the imaginary time action
\cite{mezard_variational_replica,
giamarchi_vortex_long,giamarchi_columnar_variat},
implemented by choosing a trial
action:
\begin{eqnarray}
S_{0} = \frac{1}{2 \beta \hbar} \int_q \sum_{a,b=1}^k\sum_n
G^{-1}_{ab} u_a(q,\omega_n) u_b(-q,-\omega_n)  \label{Svar}
\end{eqnarray}
where $k\to 0$ is the number of replicas. $S_0$ minimizes the
variational free energy $F_{\text{var}} = F_0 + \frac{1}{\beta
\hbar} \langle S_{\text{eff}} - S_{0} \rangle_{S_0}$, where $F_0$
denotes the free energy calculated with the trial action $S_0$.
The specific heat is defined as follows
\begin{eqnarray}
C_v(T) = -T \frac{\partial^2 F}{\partial T^2} = \frac{\partial
\overline{\langle H \rangle}}{\partial T} \label{Chalspedef}
\end{eqnarray}
where $\langle ... \rangle$ denotes a quantum and thermal average and
$\overline {...}$ disorder averaged. Introducing replicas, one finds
$\overline{\langle H \rangle}/\Omega = \frac{1}{k}\langle \sum_a
H_{\text{ph}}(u_a,\Pi_a) + \frac{2}{\beta \hbar}S_{\text{dis}}
\rangle_{S_{\text{eff}}}$. Integrating over the $\Pi_a$ fields we compute
the resulting average using $S_0$ (which amounts
exactly - thanks to the variational equations - to
compute $C_v$ using the variational free energy instead of the exact
one). We obtain:
\begin{eqnarray}
&&\overline{\langle H \rangle}/(m \Omega) = \frac{1}{\beta} \sum_{n} \int_q
\frac{1}{2} +
\frac{c}{2} (q^2 - \frac{1}{v^2} \omega_n^2) \tilde{G}(q,i\omega_n)
\nonumber \\
&&+ \frac{1}{\hbar} \int_0^{\beta \hbar} d \tau (V(\hbar \tilde{B}(\tau))
- \int_0^1 du
V(\hbar B(u)) ) \label{Hquantmoy}
\end{eqnarray}
where $\Omega$ is the volume of the system, $V(B) = - \rho_0^2
\sum_K \Delta_K \exp(- B K^2/2)$ and $\omega_n = 2\pi n/\beta
\hbar$ are the Matsubara frequencies. We have implicitly taken the
limit $k \to 0$ in (\ref{Hquantmoy}) :
we denote $\tilde{G}(q,\omega_n) = G_{aa}(q,\omega_n)$ and
parametrize $G_{a \neq b}(q,\omega_n)$ by $G(q,u)$, where $0 < u
<1$, which is $\omega_n$ independent. One has
\begin{eqnarray}
 \tilde{G}(q,\omega_n) &=& \frac{1}{c(q^2 + \omega_n^2/v^2) + \Sigma
 + I(\omega_n)} \\
 & & + \delta_{n,0} \frac{1}{cq^2}(\frac{1}{u_c}
 \frac{\Sigma}{cq^2 + \Sigma} + \int_0^{u_c} \frac{dv}{v^2}
 \frac{[\sigma](v)}{cq^2 + [\sigma](v)})\nonumber
\end{eqnarray}
Similarly we take $\hbar B_{ab}(\tau) =
\langle [u_a(x,\tau) - u_b(x,0)]^2\rangle/m$ with
$\tilde{B}(\tau)$ and $B(u)$ which is $\tau$ independent.
$u_c$ and $[\sigma](v)$ are defined in Ref. \cite{giamarchi_vortex_long,giamarchi_columnar_variat}.
$\overline{\langle H \rangle}/(m \Omega)$ is then calculated using the
solution of the variational equations
\cite{giamarchi_vortex_long,giamarchi_columnar_variat} where the
best trial Gaussian action (\ref{Svar}) is obtained by breaking
the replica-symmetry (RSB). Although the RSB scheme, and the
behavior of $B(u)$, depend in general on $d$ and on $V(B)$,
$C_v(T)$ depends on $T$ only through $u_c$, $\Sigma$ and $B = B(u > u_c)$:
\begin{eqnarray}
&& 1 = - 4V''(B)\int_q \frac{1}{(cq^2+\Sigma)^2} \label{marginality} \\
&& \Sigma = c R_c^{-2} \label{larkin_length}
\end{eqnarray}
where $R_c$ is the Larkin length, and (\ref{marginality}) is the
so called marginality condition (MC) which automatically
holds here for $d \geq 2$. We recall here the variational equation
\cite{giamarchi_columnar_variat}, relevant for the computation of
$C_v(T)$ :
\begin{equation}
I(\omega_n)=\frac{2}{\hbar} \int_0^{\beta \hbar} d\tau
(1-\cos(\omega_n \tau)) (V'(\tilde{B}(\tau)) - V'(B) ) \label{EqVariat}
\end{equation}
together with (\ref{marginality}), where
\begin{eqnarray}
 B &=& \frac{2}{\beta} \sum_n \int_q \frac{1}{cq^2 +
 c\omega_n^2/v^2 + \Sigma + I(\omega_n)} \label{matsum} \\
 \tilde{B}(\tau) &=& \frac{2}{\beta} \sum_n \int_q \frac{1
 -\cos(\omega_n\tau)}{cq^2 + c\omega_n^2/v^2 + \Sigma +
 I(\omega_n)} \nonumber
\end{eqnarray}
The solution of these coupled equations can be organized as an
expansion in powers of $\hbar$, keeping $\beta \hbar$ fixed:
$\Sigma = \Sigma_0 + \hbar \Sigma_1(\beta \hbar) + O(\hbar^2)$,
$I(\omega_n) = I_0(\omega_n) + \hbar I_1(\omega_n,\beta \hbar) +
O(\hbar^2)$ where one finds that the lowest order solution does not depend
explicitly on $\beta \hbar$ (of course $I(\omega_n)$ always
depends implicitly on $\beta \hbar$ through $\omega_n$). In
general, $\overline{\langle H \rangle}/\Omega$ and $C_v(T)$ will be
functions both
of $\hbar$ and  $\beta \hbar$ :  $ \overline{\langle H \rangle}/(m \Omega)
(\hbar,\beta
\hbar) =  H_0 + \hbar H_1(\beta \hbar) + \hbar^2 H_2(\beta
\hbar)...$, where $H_0$ is independent of $\beta \hbar$ from which
it follows that
\begin{eqnarray}
C_v(T) = C_v(\hbar,\beta \hbar) = C_0(\beta \hbar) + \hbar C_1(\beta
\hbar) + ...
\end{eqnarray}
After some manipulations, we find for $d \geq 2$
\begin{eqnarray}
\hbar H_1 = \frac{\hbar}{\beta \hbar} \sum_n \int_q
\frac{cq^2 + \Sigma_0 +
I_0(\omega_n)}{cq^2+\frac{c}{v^2}\omega_n^2+\Sigma_0+I_0(\omega_n)}
\label{Hmatsublowesthbar}
\end{eqnarray}
where in that limit $\Sigma_0$ is fixed by (\ref{marginality})
setting $B=0$ and $I_0(\omega_n)$ satisfies the equation
\cite{giamarchi_columnar_variat}
\begin{eqnarray}
&&I_0(\omega_n) = -4V''(0) (J_1(\Sigma_0) - J_1(\Sigma_0 +
\frac{c}{v^2}\omega_n^2 + I_0(\omega_n))) \nonumber \\
&&J_n(z) = \int_q \frac{1}{(cq^2+z)^n}
\end{eqnarray}
from which we can extract the low $\omega$ behavior of its
analytic continuation $I_0(\omega_n \to -i\omega + 0^+) =
I_0'(\omega) + i I_0''(\omega)$~:
\begin{eqnarray}
I_0'(\omega) \simeq A  \omega^2  \quad , \quad
I_0''(\omega) \simeq - B
 \omega
\label{SolEqVarhbar0}
\end{eqnarray}
with $A=\frac{c}{v^2}(1-\frac{J_2 J_4}{2
J_3^2})$ and $B=\sqrt{\frac{c}{v^2}\frac{J_2}{J_3}}$,
where $J_n=J_n(\Sigma_0)$. Note that (\ref{Hmatsublowesthbar})
correctly yields the equipartition for $\beta \to \infty$ :
$\hbar H_1 = T \int_q$.
The expression (\ref{Hmatsublowesthbar})
is more illuminating if we use  a spectral representation of the Green
function to transform the discrete sum over the Matsubara frequencies
in an integral :
\begin{eqnarray}
&&\hbar H_1 =  \int_{-\infty}^{+\infty} \frac{d\omega}{\pi} \hbar
\omega \rho(\omega) f_B(\omega)  \label{HFree} \\
&& \rho(\omega) = \frac{c}{v^2}\omega \int_q \text{Im} G_c(q,\omega_n
\to -i\omega + 0^+)  \\
&& = \int_q \frac{c}{v^2}\omega \frac{- I_0''(\omega)}{(cq^2 -
\frac{c}{v^2}\omega^2+\Sigma_0+I_0'(\omega))^2
+ (I_0''(\omega))^2} \label{density_of_states} \nonumber
\end{eqnarray}
where $G^{-1}_c(q,\omega_n) = \sum_b G^{-1}_{ab}(q,\omega_n)$,
 $\rho(\omega)$ is the density of states
 and $f_B(\omega)$ the Bose factor. Eq.~(\ref{HFree}) is simply the
 internal energy of free
excitations, whose density of states is given by $\rho(\omega)$,
computed self-consistently within the variational method.
As $\rho(\omega)$ does not depend,
at this order, on temperature, all the temperature dependence is
contained in the Bose factor. We thus obtain the specific heat
\begin{eqnarray}
C_v = (\beta \hbar)^2 \int_{-\infty}^{+\infty} \frac{d\omega}{4 \pi}
\frac{\rho(\omega) \omega^2}{\sinh^2{\beta \hbar \omega/2}} +
O(\hbar) \label{chal_spe_lowest_order}
\end{eqnarray}
Due to the Bose factor, the low $T$ behavior is governed by the
low $\omega$ behavior of the density of states. From
(\ref{SolEqVarhbar0}) we see that $\rho(\omega) \sim - \omega
I_0''(\omega) \sim \omega^2$, which leads to $C_v(T) \sim T^3$ in
all dimensions $d \geq 2$. Surprisingly, the linear term in the
specific heat cancels at lowest order in $\hbar$. This
cancellation, and the resulting $T^3$ behavior
occurs in fact to {\it all orders} in $\hbar$, as is
shown below. This results in the following behavior at low temperature
\footnote{The cumbersome expression for $\overline{\langle H
\rangle}/\Omega$ will be given elsewhere \cite{schehr_chalspe_long} but we
just mention that it does not reduce anymore to the ``free
particles'' contribution (\ref{HFree}).}
\begin{eqnarray}
C_v(\hbar,\beta\hbar) = (C_0  +
\hbar C_1 + \cdots) \left( \frac{T}{\hbar} \right)^3  + O((T/\hbar)^4)
\end{eqnarray}
$C_1$ will be given in \cite{schehr_chalspe_long} and we discuss
here the dominant contribution (\ref{chal_spe_lowest_order}).
Although the $T^3$ behavior is independent of the dimension $2 \leq d \leq 4$,
the coefficient itself does depend on it. More precisely
\begin{eqnarray}
&&C_v \sim \frac{8 \pi^3}{15}
\sqrt{\left(\frac{c}{v^2}\right)^3 \frac{J_2(\Sigma_0)^3}{J_3(\Sigma_0)}}
\left(\frac{T}{\hbar}\right)^3  + O(\hbar,(\beta \hbar)^{-4})
\nonumber \\
&& = \frac{4 \pi^4}{15} K_d R_c^{3-d}
F_{C_v}\left(\frac{R_c}{a}\right) \left(\frac{T}{\hbar v} \right)^3 +
 O(\hbar,(\beta \hbar)^{-4}) \nonumber \\
\end{eqnarray}
where $K_d = S_d/(2 \pi)^d$, $S_d$ being the volume of the sphere
in dimension $d$ and $F_{C_v}(x)$ a scaling function. Choosing a
spherical Brillouin zone, $F_{C_v}(x) =
2/\pi(f_2(x)^3/f_3(x))^{1/2}$, with $f_n(x) = \int_0^{2\pi x}du
u^{d-1}/(u^2+1)^n$. If we compare this result to the Debye law for
pure phonons,
\begin{eqnarray}
C_{v\text{Debye}} \sim \frac{4\pi^4}{15} K_d \left(\frac{T}{\hbar v}
\right)^d \label{Debye_law}
\end{eqnarray}
we see that $d=3$ appears as a
particular dimension below
which the specific heat is lowered by the disorder although it is
enhanced above. In three dimensions it remains of the Debye form
with a prefactor governed by the scaling function $F_{C_v}(x)$:
\begin{eqnarray}
&&F_{C_v}(x) \sim 1 \quad x \gg 1 \\
&&F_{C_v}(x) \sim \frac{16 \pi^2}{3} x^3 \quad x \ll 1
\end{eqnarray}
Note that since one does recover exactly the
Debye law at vanishing disorder, the
corrections at weak disorder are rather small.
At stronger disorder $R_c \sim a$ the
specific heat is significantly lowered by the disorder.

In the case $d < 2$, it was
noticed in \cite{giamarchi_columnar_variat} that to
obtain the correct behavior of dynamical quantities (e.g. the
conductivity $\sigma(\omega) \sim \omega^2$) through linear
response within the Matsubara equilibrium approach, one should
enforce the MC (\ref{marginality}),
which arises naturally within the dynamical methods
\cite{cugliandolo_quantum_p_spin}. If we use this condition we
obtain for $C_v(T)$ the expression (\ref{chal_spe_lowest_order})
where in the numerator $\frac{c}{v^2} \omega^2$ is replaced by
$-\frac{2-d}{2(4-d)} \Sigma_0 + \frac{c}{v^2} \omega^2 $, and this
leads to a negative specific heat at low $T$. However, if we use
the thermodynamic condition, i.e. compute $u_c$ by minimizing the
variational free energy, the internal energy can again be written as
(\ref{Hmatsublowesthbar}) which guarantees the equipartition
and a positive $C_v(T)$. Then $I(\omega_n)
\sim \omega_n^2$, and $C_v(T)$ vanishes exponentially at low $T$,
a signature of a gapped
excitation spectrum. This apparent discrepancy between
thermodynamic and dynamic quantities within the GVM, which occurs
only for $d < 2$, is a hint of
possible complications due to slow equilibration in these systems,
as observed e.g. in Coulomb glass systems
(\cite{ovadyahu_coul_glass,cohen_coul_glass}). 

Extracting the $T$-dependence to all orders in
$\hbar$ is quite difficult, but we found a 
drastic simplification of
the temperature dependence, at order $T^2$,
of $\tau$ integrals
such as in (\ref{Hquantmoy}) and (\ref{EqVariat}). Schematically,
for {\it any} polynomial form $V(x)=x^p$, expressing $B^p$ 
and $\tilde B(\tau)^p$ as multiple Matsubara sums using
(\ref{matsum}) and performing the $\tau$ integral,
one explicitly checks that cancellations occur
term by term in the multinomial expansion of $\tilde B(\tau)^p$. Details are
given in \cite{schehr_chalspe_long}. This further allows to
derive the cancellation of the contribution linear in temperature
of $C_v$ to all orders, arising naturally as
{\it a consequence of the marginality condition} (\ref{marginality}),
which, here for $d \geq 2$, holds automatically.
We find that the self-energy takes the interesting form:
\begin{eqnarray}
&& \Sigma + I(\omega_n) = \Sigma + M (1 - \delta_{n 0}) +
\tilde{I}(\omega_n) \label{sum1} \\
&& \Sigma + M = \Sigma^{(0)} + O(T^4) \label{sum2}\\
&& \tilde{I}(\omega_n) = \tilde{I}^{(0)}(\omega_n) + T^4
\tilde{I}^{(2)}(\omega_n) + \cdots
\label{sum3}
\end{eqnarray}
with $\Sigma = \Sigma^{(0)} + O(T^2)$ and
$\tilde{I}(\omega_n) \sim |\omega_n| + O(\omega_n^2)$.
The absence of $T^2$, $T^3$ dependences in Eq. (\ref{sum2},\ref{sum3})
allows to show, using marginality, that no $T^2$ nor $T^3$ term
survives in (\ref{Hquantmoy}) for $d \geq 2$. For $d<2$ the same
conclusion as above persists (simply replacing $\Sigma_0$
by $\Sigma^{(0)}$).

The drastic simplification in $T$ dependence found here
is also useful to compute analytically the putative
linear term in $C_v(T)$ in the quantum spin glasses. Both
models in \cite{georges_mf_quantum_spinglass} and
\cite{cugliandolo_quantum_p_spin} are solved via a
one step RSB solution, hence analogous to our $d<2$ case, and thus
are gapless provided the marginality condition is enforced (see
discussion above). Ref. \cite{cugliandolo_quantum_p_spin} studies the
specific heat of the quantum $p$-spin model and obtain
numerical evidence for a linear contribution. We find that
this term vanishes.
In Ref. \cite{georges_mf_quantum_spinglass} the SU(N)
fully connected Heisenberg spin glass (of spin $S$) was solved for infinite
$N$. There too it was found that the dominant term (analogous to $C_0$
here) in a $1/S$ semiclassical expansion also behaves as $\sim
T^3$. Whether it holds to higher order in $1/S$ has
been discussed in \cite{parcollet_private_chaleur_spe}
and \cite{rozenberg_t2} and is currently under investigation
\cite{schehr_chalspe_long}. In all cases a structure similar
to (\ref{sum1},\ref{sum2},\ref{sum3}) also holds.

One can discuss the validity of the mean field method.
The GVM can also be used to describe non
periodic systems (e.g. directed polymer) in terms of a power law
$V(B)$ \cite{mezard_variational_replica,giamarchi_vortex_long} and
then becomes exact for $m \to \infty$
with $R(u) = - m V(u^2/m)$. We thus expect that the $T^3$ behavior
is exact in this limit in the full RSB case. For the periodic problem it is always an
approximation which describes the vibrational phonon-like
excitations of the pinned system and may not treat
accurately excitations such as solitons. Whether these solitons
or other type of excitations such as
dislocations for a lattice could reestablish a linear contribution
to the specific heat from two level systems type behavior remains
to be investigated.

This weak T-dependence of the specific heat in the quantum regime is
relevant to the so-called vortons modes of the vortex
lattice of superconductors. Other contributions
such as core electrons \cite{caroli_core_states}, or their interaction
with vortons (responsible for the dissipation in the quantum
dynamics of the vortons) yield linear specific heat
\cite{blatter_chalspe} which may dominate over the present one
in the quantum regime. Whether this also holds in the classical
regime for the vortons is examined in
\cite{schehr_chalspe_classique}.

In conclusion we have obtained from first
principles the specific heat
of an elastic quantum pinned system, such as the
Bragg glass phase of a lattice in presence of
impurities. We found that $C_v \sim T^3$
in $d=2,3$ due to a non trivial mechanism of
cancellation of the linear term. This simplification
which occurs from the marginality condition
will be useful for a general understanding of
the specific heat behavior within the
mean field theory.
The question remains concerning the contribution of other
types of excitations such as plastic deformations.
Further analytical and numerical investigations
would be desirable to a better understanding of the
behavior of the specific heat in glasses.

\begin{acknowledgments}
We acknowledge G. Blatter, L. F. Cugliandolo, S. Florens, A. Georges,
L. Ioffe, T. Klein, P. Monceau, O. Parcollet and  C. Varma for
helpful discussions.
\end{acknowledgments}


\begin{thebibliography}{32}
\expandafter\ifx\csname natexlab\endcsname\relax\def\natexlab#1{#1}\fi
\expandafter\ifx\csname bibnamefont\endcsname\relax
  \def\bibnamefont#1{#1}\fi
\expandafter\ifx\csname bibfnamefont\endcsname\relax
  \def\bibfnamefont#1{#1}\fi
\expandafter\ifx\csname citenamefont\endcsname\relax
  \def\citenamefont#1{#1}\fi
\expandafter\ifx\csname url\endcsname\relax
  \def\url#1{\texttt{#1}}\fi
\expandafter\ifx\csname urlprefix\endcsname\relax\def\urlprefix{URL }\fi
\providecommand{\bibinfo}[2]{#2}
\providecommand{\eprint}[2][]{\url{#2}}

\bibitem[{\citenamefont{Zeller and Pohl}(1971)}]{zeller_chalspe_struct_glasses}
\bibinfo{author}{\bibfnamefont{R.~C.} \bibnamefont{Zeller}} \bibnamefont{and}
  \bibinfo{author}{\bibfnamefont{R.~O.} \bibnamefont{Pohl}},
  \bibinfo{journal}{Phys. Rev. B} \textbf{\bibinfo{volume}{4}},
  \bibinfo{pages}{2029} (\bibinfo{year}{1971}).

\bibitem[{bin()}]{binder_spinglass_review}
\bibinfo{note}{K. Binder, A. P. Young, Rev. Mod.
  Phys. {\bf 58} 801 (1986)}.



\bibitem[{\citenamefont{Phillips}(1978)}]{phillips_chalspe_amorphous}
\bibinfo{author}{\bibfnamefont{W.~A.} \bibnamefont{Phillips}},
  \bibinfo{journal}{J. Non-Crys. Solids} \textbf{\bibinfo{volume}{31}},
  \bibinfo{pages}{267} (\bibinfo{year}{1978}).

\bibitem[{\citenamefont{Ackerman et~al.}(1981)\citenamefont{Ackerman, Moy,
  Potter, and Anderson}}]{ackerman_chalspe_disordered_crystals}
\bibinfo{author}{\bibfnamefont{D.~A.} \bibnamefont{Ackerman et~al.}},
   \bibinfo{journal}{Phys. Rev. B}
  \textbf{\bibinfo{volume}{23}}, \bibinfo{pages}{3886} (\bibinfo{year}{1981}).


\bibitem[{\citenamefont{Anderson et~al.}(1972)\citenamefont{Anderson, Halperin,
  and Varma}}]{anderson_twolevels}
\bibinfo{author}{\bibfnamefont{P.~W.} \bibnamefont{Anderson}},
  \bibinfo{author}{\bibfnamefont{B.~I.} \bibnamefont{Halperin}},
  \bibnamefont{and} \bibinfo{author}{\bibfnamefont{C.~M.} \bibnamefont{Varma}},
  \bibinfo{journal}{Phil. Mag. B} \textbf{\bibinfo{volume}{25}},
  \bibinfo{pages}{1} (\bibinfo{year}{1972}).






\bibitem[{\citenamefont{Dianoux
et~al. and~ref.~therein}(1993)}]{boson_peak_review}
\bibinfo{author}{\bibfnamefont{A.J.}~\bibnamefont{Dianoux et.~al}} in
  \emph{\bibinfo{booktitle}{Dynamics of Disordered Materials II.}},
  (\bibinfo{publisher}{North-Holland}, \bibinfo{address}{Amsterdam},
  \bibinfo{year}{1993}) and~Ref.~therein.









\bibitem[{\citenamefont{Schirmacher et~al. and ref therein}(2002)\citenamefont{Schirmacher
et~al.}}]{schirmacher_bosonpeak}
\bibinfo{author}{\bibfnamefont{W.} \bibnamefont{Schirmacher et~al.}},
  \bibinfo{journal}{Phys. Rev. Lett.} \textbf{\bibinfo{volume}{81}},
  \bibinfo{pages}{136} (\bibinfo{year}{1998}).

\bibitem[{\citenamefont{Grigera
et~al.}(2001)\citenamefont{Grigera et~al.}}]{grigera_vibrational_spectrum}
\bibinfo{author}{\bibfnamefont{T.~S.} \bibnamefont{Grigera et al.}},
  \bibinfo{journal}{Phys. Rev. Lett.} \textbf{\bibinfo{volume}{87}},
  \bibinfo{pages}{85502} (\bibinfo{year}{2001}).


\bibitem[{\citenamefont{Schehr et~al.}(2003)\citenamefont{Schehr, Giamarchi,
  and Doussal}}]{schehr_chalspe_classique}
\bibinfo{author}{\bibfnamefont{G.}~\bibnamefont{Schehr et~al.}},
  (\bibinfo{year}{2003}), \bibinfo{note}{cond-mat/0301053}.




\bibitem[{\citenamefont{Schirmacher et~al.}(2002)\citenamefont{Schirmacher
et~al.}}]{schirmacher_anharmonic}
\bibinfo{author}{\bibfnamefont{W.} \bibnamefont{Schirmacher et~al.}},
  \bibinfo{journal}{Phys. Stat. Sol. B} \textbf{\bibinfo{volume}{230}},
  \bibinfo{pages}{31} (\bibinfo{year}{2002}).




\bibitem[{\citenamefont{Giamarchi and {Le
  Doussal}}(1996)}]{giamarchi_columnar_variat}
\bibinfo{author}{\bibfnamefont{T.}~\bibnamefont{Giamarchi}} \bibnamefont{and}
  \bibinfo{author}{\bibfnamefont{P.}~\bibnamefont{{Le Doussal}}},
  \bibinfo{journal}{Phys. Rev. B} \textbf{\bibinfo{volume}{53}},
  \bibinfo{pages}{15206} (\bibinfo{year}{1996}).

\bibitem[{\citenamefont{M{\'e}zard and
  Parisi}(1991)}]{mezard_variational_replica}
\bibinfo{author}{\bibfnamefont{M.}~\bibnamefont{M{\'e}zard}} \bibnamefont{and}
  \bibinfo{author}{\bibfnamefont{G.}~\bibnamefont{Parisi}},
  \bibinfo{journal}{J. de Phys. I} \textbf{\bibinfo{volume}{1}},
  \bibinfo{pages}{809} (\bibinfo{year}{1991}).


\bibitem[{\citenamefont{Georges et~al.}(2001)\citenamefont{Georges, Parcollet,
  and Sachdev}}]{georges_mf_quantum_spinglass}
\bibinfo{author}{\bibfnamefont{A.}~\bibnamefont{Georges}},
  \bibinfo{author}{\bibfnamefont{O.}~\bibnamefont{Parcollet}},
  \bibnamefont{and} \bibinfo{author}{\bibfnamefont{S.}~\bibnamefont{Sachdev}},
  \bibinfo{journal}{Phys. Rev. B} \textbf{\bibinfo{volume}{63}},
  \bibinfo{pages}{134406} (\bibinfo{year}{2001}).

\bibitem[{\citenamefont{Cugliandolo et~al.}(2001)\citenamefont{Cugliandolo,
  Grempel, and {da Silva Santos}}}]{cugliandolo_quantum_p_spin}
\bibinfo{author}{\bibfnamefont{L.~F.} \bibnamefont{Cugliandolo et~al.}},
   \bibinfo{journal}{Phys. Rev. B}
  \textbf{\bibinfo{volume}{64}}, \bibinfo{pages}{14403} (\bibinfo{year}{2001}).


\bibitem[{\citenamefont{Camjayi and {Rozenberg}}(2002)}]{rozenberg_t2}
\bibinfo{author}{\bibfnamefont{A.}~\bibnamefont{Camjayi}} \bibnamefont{and}
  \bibinfo{author}{\bibfnamefont{M.~J.}~\bibnamefont{{Rozenberg}}},
  \bibinfo{journal}{cond-mat/0210407} \textbf{\bibinfo{volume}{}},
  \bibinfo{pages}{} (\bibinfo{year}{2002}).


\bibitem[{\citenamefont{Blatter et~al.}(1994)\citenamefont{Blatter, Feigel'man,
  Geshkenbein, Larkin, and Vinokur}}]{blatter_vortex_review}
\bibinfo{author}{\bibfnamefont{G.}~\bibnamefont{Blatter et~al.}},
\bibinfo{journal}{Rev. Mod. Phys.}
  \textbf{\bibinfo{volume}{66}}, \bibinfo{pages}{1125} (\bibinfo{year}{1994}).



\bibitem[{\citenamefont{Giamarchi and {Le
  Doussal}}(1998)}]{giamarchi_book_young}
\bibinfo{author}{\bibfnamefont{T.}~\bibnamefont{Giamarchi}} \bibnamefont{and}
  \bibinfo{author}{\bibfnamefont{P.}~\bibnamefont{{Le Doussal}}},
  \emph{\bibinfo{title}{Statics and dynamics of disordered elastic systems}}
  (\bibinfo{publisher}{World Scientific}, \bibinfo{address}{Singapore},
  \bibinfo{year}{1998}), p. \bibinfo{pages}{321},
  \bibinfo{note}{cond-mat/9705096}.

\bibitem[{\citenamefont{Nattermann and
  Scheidl}(2000)}]{nattermann_vortex_review}
\bibinfo{author}{\bibfnamefont{T.}~\bibnamefont{Nattermann}} \bibnamefont{and}
  \bibinfo{author}{\bibfnamefont{S.}~\bibnamefont{Scheidl}},
  \bibinfo{journal}{Adv. Phys.} \textbf{\bibinfo{volume}{49}},
  \bibinfo{pages}{607} (\bibinfo{year}{2000}).


\bibitem[{\citenamefont{Giamarchi and
  Bhattacharya}(2002)}]{giamarchi_vortex_review}
\bibinfo{author}{\bibfnamefont{T.}~\bibnamefont{Giamarchi}} \bibnamefont{and}
  \bibinfo{author}{\bibfnamefont{S.}~\bibnamefont{Bhattacharya}}, in
  \emph{\bibinfo{booktitle}{High Magnetic Fields: Applications in Condensed
  Matter Physics and Spectroscopy}}, edited by
  \bibinfo{editor}{\bibfnamefont{C.}~\bibnamefont{{Berthier et al.}}}
  (\bibinfo{publisher}{Springer-Verlag}, \bibinfo{address}{Berlin},
  \bibinfo{year}{2002}),
  \bibinfo{note}{cond-mat/0111052}.

\bibitem[{\citenamefont{Giamarchi}(2002)}]{giamarchi_wigner_review}
\bibinfo{author}{\bibfnamefont{T.}~\bibnamefont{Giamarchi}}, in
  \emph{\bibinfo{booktitle}{Strongly correlated fermions and bosons in low
  dimensional disordered systems}}, edited by
  \bibinfo{editor}{\bibfnamefont{I.~V.} \bibnamefont{{Lerner et al.}}}
  (\bibinfo{publisher}{Kluwer}, \bibinfo{address}{Dordrecht},
  \bibinfo{year}{2002}), \bibinfo{note}{cond-mat/0205099}.

\bibitem[{\citenamefont{Chitra et~al.}(2001)\citenamefont{Chitra, Giamarchi,
  and {Le Doussal}}}]{chitra_wigner_long}
\bibinfo{author}{\bibfnamefont{R.}~\bibnamefont{Chitra et~al.}},
\bibinfo{journal}{Phys. Rev. B} \textbf{\bibinfo{volume}{65}},
  \bibinfo{pages}{035312} (\bibinfo{year}{2001}).

\bibitem[{\citenamefont{Gr{\"u}ner}(1994)}]{gruner_book_cdw}
\bibinfo{author}{\bibfnamefont{G.}~\bibnamefont{Gr{\"u}ner}},
  \emph{\bibinfo{title}{Density Waves in Solids}}
  (\bibinfo{publisher}{Addison-Wesley, Reading}, \bibinfo{address}{MA},
  \bibinfo{year}{1994}).

\bibitem[{\citenamefont{Saunders et~al.}(2000)\citenamefont{Saunders,
  Radzihovsky, and Toner}}]{saunders_smectic_bragg_glass}
\bibinfo{author}{\bibfnamefont{K.}~\bibnamefont{Saunders et~al.}},
  \bibinfo{journal}{Phys. Rev. Lett.} \textbf{\bibinfo{volume}{85}},
  \bibinfo{pages}{4309} (\bibinfo{year}{2000}).


\bibitem[{\citenamefont{Caroli et~al.}(1964)\citenamefont{Caroli, {de Gennes},
  and Matricon}}]{caroli_core_states}
\bibinfo{author}{\bibfnamefont{C.}~\bibnamefont{Caroli et al.}},
  \bibinfo{journal}{Phys. Lett.} \textbf{\bibinfo{volume}{9}},
  \bibinfo{pages}{307} (\bibinfo{year}{1964}).


\bibitem[{\citenamefont{Volovik}(1993)}]{volovik_chalspe_dwaves}
\bibinfo{author}{\bibfnamefont{G.~E.} \bibnamefont{Volovik}},
  \bibinfo{journal}{JETP Lett.} \textbf{\bibinfo{volume}{58}},
  \bibinfo{pages}{469} (\bibinfo{year}{1993}).

\bibitem[{\citenamefont{Blatter and Ivlev}(1994)}]{blatter_chalspe}
\bibinfo{author}{\bibfnamefont{G.}~\bibnamefont{Blatter}} \bibnamefont{and}
  \bibinfo{author}{\bibfnamefont{B.~I.} \bibnamefont{Ivlev}},
  \bibinfo{journal}{Phys. Rev. B} \textbf{\bibinfo{volume}{50}},
  \bibinfo{pages}{10272} (\bibinfo{year}{1994}).

\bibitem[{\citenamefont{Bulaevskii and Maley}(1993)}]{bulaevskii_chalspe}
\bibinfo{author}{\bibfnamefont{L.~N.} \bibnamefont{Bulaevskii}}
  \bibnamefont{and} \bibinfo{author}{\bibfnamefont{M.~P.} \bibnamefont{Maley}},
  \bibinfo{journal}{Phys. Rev. Lett.} \textbf{\bibinfo{volume}{71}},
  \bibinfo{pages}{3541} (\bibinfo{year}{1993}).

\bibitem[{\citenamefont{Izawa et~al.}(2001)\citenamefont{Izawa, Shibata,
  Matsuda, Kato, Takeya, Hirata, {van der Beek}, and
  Konczykowski}}]{izawa_chalspe_borocabide}
\bibinfo{author}{\bibfnamefont{K.}~\bibnamefont{Izawa et~al.}},
  \bibinfo{journal}{Phys. Rev. Lett.} \textbf{\bibinfo{volume}{86}},
  \bibinfo{pages}{1327} (\bibinfo{year}{2001}).



\bibitem[{\citenamefont{Odin et~al.}(2001)\citenamefont{Odin, Lasjaunias,
  Biljakovic, Hasselbach, and Monceau}}]{odin_chalspe_blue_bronze}
\bibinfo{author}{\bibfnamefont{J.}~\bibnamefont{Odin et~al.}},
  \bibinfo{journal}{Eur. Phys. J. B} \textbf{\bibinfo{volume}{24}},
  \bibinfo{pages}{315} (\bibinfo{year}{2001}).

\bibitem[{\citenamefont{Lasjaunias et~al.}(1996)\citenamefont{Lasjaunias,
  Biljakovic, and Monceau}}]{lasjaunias_chalspe_noneq_sdw}
\bibinfo{author}{\bibfnamefont{J.~C.} \bibnamefont{Lasjaunias}},
  \bibinfo{author}{\bibfnamefont{K.}~\bibnamefont{Biljakovic}},
  \bibnamefont{and} \bibinfo{author}{\bibfnamefont{P.}~\bibnamefont{Monceau}},
  \bibinfo{journal}{Phys. Rev. B} \textbf{\bibinfo{volume}{53}},
  \bibinfo{pages}{7699} (\bibinfo{year}{1996}).





\bibitem[{\citenamefont{Giamarchi and {Le
  Doussal}}(1995)}]{giamarchi_vortex_long}
\bibinfo{author}{\bibfnamefont{T.}~\bibnamefont{Giamarchi}} \bibnamefont{and}
  \bibinfo{author}{\bibfnamefont{P.}~\bibnamefont{{Le Doussal}}},
  \bibinfo{journal}{Phys. Rev. B} \textbf{\bibinfo{volume}{52}},
  \bibinfo{pages}{1242} (\bibinfo{year}{1995}).



\bibitem[{\citenamefont{Ovadyahu}(1986)}]{ovadyahu_coul_glass}
\bibinfo{author}{\bibfnamefont{Z.}~\bibnamefont{Ovadyahu}} 
  \bibinfo{journal}{J. Phys. C: Solid State Phys.}
\textbf{\bibinfo{volume}{19}}, 
  \bibinfo{pages}{5187} (\bibinfo{year}{1986}).


\bibitem[{\citenamefont{Cohen}(1994)}]{cohen_coul_glass}
\bibinfo{author}{\bibfnamefont{O.}~\bibnamefont{Cohen~et~al.}} 
  \bibinfo{journal}{Phys. Rev. B }
\textbf{\bibinfo{volume}{50}}, 
  \bibinfo{pages}{10442} (\bibinfo{year}{1994}).


\bibitem[{\citenamefont{Schehr et~al.}(2002)\citenamefont{Schehr, Giamarchi,
  and Doussal}}]{schehr_chalspe_long}
\bibinfo{author}{\bibfnamefont{G.}~\bibnamefont{Schehr et~al.}},
  (\bibinfo{year}{2002}), \bibinfo{note}{in
  preparation.}

\bibitem[{\citenamefont{Parcollet}(2002)}]{parcollet_private_chaleur_spe}
\bibinfo{author}{\bibfnamefont{O.}~\bibnamefont{Parcollet}}
  (\bibinfo{year}{2002}), \bibinfo{note}{private communication.}










\end{thebibliography}

\end{document}